\documentclass[aps,prb,twocolumn,showpacs]{revtex4}%
\usepackage{amsfonts}
\usepackage{amsmath}
\usepackage{amssymb}
\usepackage{graphicx}
\usepackage{dcolumn}
\usepackage{bm}

\begin{document}
\title{Pair-breaking in iron-pnictides}
\author{V. G. Kogan}
\affiliation{Ames Laboratory and Department of Physics \& Astronomy, 
Iowa State University,
Ames, Iowa 50011}



\begin{abstract}
The puzzling features of the slopes of the upper critical 
field at the critical temperature $T_c$, $H^\prime_{c2}(T_c)\propto
T_c$, and of the specific heat jump
$\Delta C\propto T_c^3$  of iron-pnictides  are
interpreted as caused by a strong pair-breaking.  
\end{abstract}

\date{\today}
\maketitle

\section{Introduction}

Newly discovered iron-pnictide superconductors have a number 
of uncommon properties. The subject of this paper are two such
properties: (a) The specific heat jump $\Delta C$ 
is proportional to $T_c^3$ as demonstrated on 
``122" series of Ba(Fe$_{1-x}$Co$_x)_2$As$_2$   and  
Ba(Fe$_{1-x}$Ni$_x)_2$As$_2$.\cite{BNC}  This behavior, according
to Ref.\,\onlinecite{Z}, cannot be explained within the ``realm of
conventional BCS theory".  Similar behavior is recorded in 122
crystals with Ba substituted partially with K and with Fe substituted with 
Pd, Rh,\cite{122K} and Co-Cu.  (b)
Slopes of the upper critical field $dH_{c2}/dT$ at $T_c$ are 
proportional to $T_c$ across both 1111 and 122 series. 

It is shown below that
both scalings can be understood within the weak-coupling BCS model
provided a strong pair breaking is present  in these materials. In
fact, these features should also be present in conventional
superconductors with magnetic impurities as discussed by 
Abrikosov and Gor'kov (AG) in their  
seminal work on the pair breaking  for the nearly critical
concentration of these impurities when $T_c\ll T_{c0}$, the
critical temperature of clean material.\cite{AG} AG had considered 
  isotropic materials with a spherical Fermi surface and the
s-wave order parameter constant along this surface. The
symmetry of the order parameter in  multi-band pnictides is not yet
determined with certainty; however, many favor the $\pm s$
structure.\cite{Mazin-Schm,Junhwa} The critical temperature in materials with a strongly anisotropic order parameter is suppressed not only by  scattering breaking the time reversal symmetry (e.g., the spin-flip); in fact, any scattering reduces $T_c$. The term ``pair-breaking" is used here in a broad sense for any process suppressing $T_c$. It
is shown below that both features, $dH_{c2}/dT\propto T_c$ and  $\Delta C\propto T_c^3$, follow from the assumption that the ``pair-breaking  in a broad sense" is strong. 

Below, the linearized Ginzburg-Landau
(GL) equation and the energy near $T_c$ are derived within the weak
coupling scheme (that allows one to evaluate $dH_{c2}/dT$ and
$\Delta C$ at $T_c$) for an arbitrary anisotropy of the order parameter
$\Delta$ and of the Fermi surface  in the presence of pair-breaking.
The text is focussed on the situation when the average $\langle
\Delta\rangle$ over the F-surface is close to zero that
presumably is the case of pnictides.\cite{Mazin-Schm,Junhwa}  Comparison with the data
available  concludes the paper.
 
Perhaps, the simplest for our purpose is  
the Eilenberger quasiclassical formulation of the Gor'kov's theory  that holds for a general
anisotropic  F-surface  and for any gap symmetry:\cite{E}   
\begin{eqnarray}
{\bm v} {\bm  \Pi}f&=&2  \Delta g   -2\omega f +
\frac{g}{\tau_-}\langle f\rangle -
\frac{f}{\tau_+}\langle g\rangle \,,\label{Eil1}\\ 
 -{\bm v} {\bm  \Pi}^*f^+&=&2  \Delta^* g   -2\omega f^+ + 
\frac{g}{\tau_-}\langle f^+\rangle -
\frac{f^+}{\tau_+}\langle g\rangle \,,\label{Eil2}\\ 
g^2&=&1-ff^{+}\,, \label{Eil3} \\
 \Delta({\bm  r},{\bm k}_F)&=&2\pi TN(0) \sum_{\omega >0}^{\omega_D}
 \Big\langle V({\bm k}_F,{\bm k}_F^{\prime\,}) f({\bm v}^{\prime},{\bm 
 r},\omega)\Big\rangle_{{\bm k}_F^{\prime\,}}\quad  \label{eil4} 
\end{eqnarray}
Here, ${\bm v}$ is the Fermi velocity, ${\bm  \Pi} =\nabla +2\pi i{\bm 
A}/\phi_0$, $\phi_0$ is the flux quantum. $ \Delta ({\bm r},{\bm k}_F)$ is
 the order parameter that in general depends on the position
${\bm  k}_F$ at the  F-surface  of 
other than the isotropic s-wave symmetry.  The functions
$f({\bm  r},{\bm v},\omega),\,\,  f^{+} $, and $g$ originate from Gor'kov's
Green's functions integrated over the energy variable near the F-surface.
Further,
$N(0)$ is the total density of states at the Fermi level per one spin;
the  Matsubara frequencies are $\omega=\pi T(2n+1)$
with an integer $n$ and $\hbar=k_B=1$. The averages over
the F-surface are shown as $ \langle... \rangle$.  

  The  scattering in the Born approximation is characterized  by two
scattering times, the transport scattering time $\tau $   responsible for
conductivity in the normal state, and  $\tau_m$ for  spin-flip processes: 
  \begin{equation}
\frac{1}{\tau_\pm}=\frac{1}{\tau }\pm\frac{1}{\tau_m} \,,
  \label{taus}
  \end{equation}
The strong
scattering in   unitary limit is not considered here.
Commonly, the scattering is characterized by two parameters
  \begin{equation}
\rho=\frac{1}{2\pi T_c\tau }\qquad {\rm and}\qquad
\rho_m=\frac{1}{2\pi  T_c\tau_m
}\,,
  \label{rhos}
  \end{equation}
or equivalently by
$\rho_\pm= \rho\pm \rho_m$.
This is of course  
 a simplification;  for multi-band F-surfaces one may need 
more parameters for various intra- and inter-band processes. This
and other simplifying assumptions notwithstanding, the model
employed is amenable for analytic work and may prove useful. 

Long experience in dealing with
pair-breaking effects has shown that the formal AG scheme in fact
applies to various situations with different causes for the
pair breaking, not necessarily the AG spin-flip
scattering.\cite{Maki} In each particular situation, the  parameter 
  $\rho_m$ must be properly defined. Without
specifying the pair breaking  mechanism in materials of interest here
we apply below the AG approach to show that the pair breaking
accounts for experimental data on slopes of  $H_{c2}$ at $T_c$ and for quite unusual dependence of the
specific heat jump on $T_c$.

Commonly, the effective coupling $V$   is assumed factorizable,
 $ V({\bm k_F},{\bm k_F}^{\prime\,})=V_0 \,\Omega({\bm
k_F})\,\Omega({\bm k_F}^{\prime\,})$.\cite{Kad}  One then looks for
the order parameter in the form:   
  \begin{equation}
  \Delta (
{\bm  r},T;{\bm k_F})=\Psi ({\bm  r},T)\, \Omega({\bm k_F}) \,.
\label{D=PsiO}
\end{equation} 
 Our notation is motivated by the fact that  so defined $\Psi ({\bm  r},T)$ enters the
Ginzburg-Landau (GL) theory near   $T_c$. The  function  $\Omega({\bm  k}_F)$, which describes the variation of
$\Delta$ along the F-surface, is conveniently normalized: 
\begin{equation}
       \Big\langle \Omega^2 \Big\rangle=1\,.
\label{norm}
\end{equation}
Then, the self-consistency equation (\ref{eil4}) takes the form:
       \begin{equation}
\Psi( {\bm  r},T)=2\pi T N(0)V_0 \sum_{\omega >0}^{\omega_D} \Big\langle
\Omega({\bm k_F} ) f({\bm k_F} ,{\bm  r},\omega)\Big\rangle \,.
\label{gap}
\end{equation}
The assumption of a factorizable 
potential is quite restrictive as far as complicated F-surfaces
and interactions are concerned. E.g., within a two-band scheme with
four independent coupling constants $V_{ij}$, the factorizable
model implies  $V_{11}V_{22}-V_{12}V_{21}=0$.

Instead of dealing with the effective microscopic electron-electron
interaction $V$ and with the energy scale $\omega_D$, one can use within
the weak coupling scheme the critical temperature $T_{c0}$ (of the
hypothetic clean material) utilizing the identity
\begin{equation}
\frac{ 1}{N(0) V_0}= \ln \frac{T}{T_{c0}}+2\pi T\sum_{\omega >0}^{\omega_D}
       \frac{ 1}{ \omega} \,,
\label{1/NV}
\end{equation}
which is equivalent to the BCS relation $\Delta_0(0)=\pi T_{c0}
e^{-\gamma}=2 \omega_D
\exp(-1/N(0)V_0)$; $\gamma$ is the Euler constant.
Substitute Eq.\,(\ref{1/NV}) in
Eq.\,(\ref{gap}) and replace
$\omega_D$ with infinity due to  fast convergence:
       \begin{equation}
\frac{\Psi }{2\pi T} \ln \frac{T_{c0}}{T}= \sum_{\omega
>0}^{\infty}\left(\frac{\Psi}{ \omega}-\Big\langle \Omega \, f
\Big\rangle\right)\,.
\label{gap1}
\end{equation}

\section{GL domain and ${\bm T_c(\tau,\tau_m)}$}

Near $T_c$, $g=1-ff^{+}/2$ and Eq.\,(\ref{Eil1}) reads:
\begin{eqnarray}
\frac{1}{2}{\bm v} {\bm  \Pi}f&=&  \Delta  -\omega_+ f    +
\frac{\langle f\rangle}{2\tau_-}\nonumber\\
&-&\frac{ff^{+}}{2}\left(\Delta+\frac{\langle f\rangle}{2\tau_-}\right)+\frac{f\langle ff^{+}\rangle}{2\tau_+}
 \,.\label{Eil11}
\end{eqnarray}
Here, 
\begin{equation}
\omega_+=\omega + 1/2\tau_+  \,,
  \label{w+}
 \end{equation}
and the terms on the RHS are arranged according to the their 
order in powers of $\delta t=1-T/T_c$: the terms on the upper line
are of the order $\delta t^{1/2}$ whereas on the lower line
$\sim\delta t^{3/2}$. Note  that on the LHS, $\Pi f\sim f/\xi\sim
\delta t$. 

We look for the solution $f=f_1+f_2+...$ where $ f_1\sim\delta t^{1/2}$ and $ f_2\sim\delta t  $. Hence, we have in the lowest order:
\begin{eqnarray}
0=  \Delta  -\omega ^+ f_1    +
\frac{\langle f_1\rangle}{2\tau_-}  \,.\label{eqf1}
\end{eqnarray}
Taking the average over the Fermi surface we obtain
\begin{eqnarray}
\langle f_1\rangle =   \langle  \Delta\rangle /\omega_{ m}   \,,\qquad\omega_{ m}=\omega  +1/\tau_m   \label{<f1>}
\end{eqnarray}
(note the difference in definitions of $\omega_+$ and 
$\omega_m$). Hence:
\begin{eqnarray}
  f_1 =\frac{1}{\omega ^+}\left(\Delta + \frac{ \langle  
\Delta\rangle }{2\tau_-\omega_{ m}}\right)  \, . \label{f_1}
\end{eqnarray}

Comparing terms of the order $\delta t$, we get
\begin{eqnarray}
\langle f_2\rangle =  -\frac{\langle {\bm v} {\bm  \Pi}f_1\rangle}{2\omega_{ m}}=0  \,,   \label{<f2>}
\end{eqnarray}
and 
\begin{eqnarray}
  f_2 =-\frac{1}{2\omega ^+\omega_{ m}} {\bm v} {\bm\Pi}\left(\Delta
+ \frac{ \langle  \Delta\rangle }{2\tau_-\omega_{ m}}\right)  \, .
\label{f_2}
\end{eqnarray}
 
Evaluation of higher order corrections for arbitrary $\Delta$
anisotropy is increasingly cumbersome unlike the case $\langle 
\Delta\rangle =0$ for which one finds:
\begin{eqnarray}
f_3=-\frac{\Delta}{2 \omega_+^2\omega_m}\left(\Delta^2-\frac{
\langle\Delta^2\rangle}{2\tau_+ \omega ^+}\right)\,.  
\label{f_3}
\end{eqnarray}


The critical temperature of materials with anisotropic order parameter is suppressed   by   
scattering.     In zero field,   all quantities are coordinate independent; besides, as
$T\to T_c$, $g \to 1$.  Therefore, we can utilize $f$ of 
Eq.\,(\ref{f_1}) in the lowest order to obtain for $T_c$:
      \begin{equation}
\frac{1}{2\pi T_c} \ln \frac{T_{c0}}{T_c}= \sum_{\omega
>0}^{\infty}\Big(\frac{1}{ \omega_c}-\frac{1}{ \omega^+_c}-
\frac{\langle\Omega  \rangle^2}{2 \omega^m_c\omega_c^+\tau_-}\Big)\,,
\label{Tc}
\end{equation}
where the subscript $c$ is to denote that  $\omega$'s 
 are taken at $T_c$. This generalization of the well-known
AG result   gives the $T_c$
suppression for any (Born) scattering for arbitrary symmetry of the
order parameter; it has originally been  obtained by
Openov.\cite{Openov} 
The sums here are expressed in terms of di-gamma functions:  
\begin{eqnarray}
\ln
\frac{T_{c0}}{T_c}&=&\psi\left(\frac{1+\rho^+}{2}\right)-
\psi\left(\frac{1}{2}\right)\nonumber\\
&-&\langle \Omega\rangle^2
\left[\psi\left(\frac{1+\rho^+}{2}\right)-
\psi\left(\frac{1}{2}+\rho_m\right)\right].\quad
\label{result}
\end{eqnarray}

If $T_c\to 0$, one can use asymptotic expansion   $\psi(z)=\ln\,z-1/2z$ for
large arguments since
$\rho,\rho_m\to\infty$. The leading term then gives  that $T_c= 0$
when scattering times satisfy the relation:
\begin{eqnarray}
\frac{1}{\tau_m}\left(\frac{\tau_m}{2\tau^+}\right)^{1-\langle
\Omega\rangle^2} =\frac{\Delta_0(0)}{2 }\,.
\label{crit}
\end{eqnarray}
Here, $\Delta_0(0)=\pi T_{c0}\,e^{-\gamma}$ is the zero temperature gap of the (hypothetic) scattering free material. Clearly, this  reduces to the AG critical rate
$1/\tau_m=\Delta_0(0)/2 $  for isotropic order parameters. If $\langle\Omega\rangle =0$ (as, e.g., for the d-wave), we have the critical combined rate:
$ 1/\tau^+ = \Delta_0(0) $.

      In the absence of spin-flip scattering   
($\tau_m\to\infty$) the LHS is zero and Eq.\,(\ref{crit}) has
no solutions for $\tau$, i.e., $T_c$  does not turn zero for any $\tau$.
 However, a finite $\tau$ at which
$T_c=0$ does exists for any finite $\tau_m$. One can show that near the critical value $\tau_{+,crit}^{ \langle\Omega\rangle^2-1}=\Delta_0(0)(\tau_m/2)^{ \langle
\Omega\rangle^2} $, the critical temperature behaves similarly to the AG gapless case, $T_c\propto(\tau_+ -\tau^+_{crit})^{1/2}$.

Combining Eqs.\,(\ref{gap1}) and (\ref{Tc})
one excludes the unphysical $T_{c0}$:
      \begin{equation}
\frac{\Psi}{2\pi T } \ln \frac{T_{c}}{T }= \sum_{\omega
>0}^{\infty}\left(\frac{\Psi}{t \omega_c^+} +\frac{\Psi\langle\Omega  \rangle^2}{2t \omega^m_c\omega_c^+\tau_-}-\Big\langle \Omega \, f
\Big\rangle\right)  \, 
\label{self-cons-a}
\end{equation} 
where $t=T/T_c$.

\section{The case $\bm T_c\ll \bm T_{c0}$}

  Situations of interested here are of $T_c$   strongly 
suppressed relative to $T_{c0}$ (similar to the gapless
superconductivity of AG, but not necessarily the same). 
It is convenient for this
purpose to rearrange Eq.\,(\ref{self-cons-a}) by adding 
and subtracting $\Psi/ \omega ^+$ under the sum. We transform:
      \begin{eqnarray}
&&2\pi T  \sum_{\omega
 0}^{\infty}\left(\frac{1}{t \omega_c^+} -\frac{1}{ \omega ^+ }\right) \nonumber\\
&& = \sum_{n=
>0}^{\infty}
 \left(\frac{1}{n+1/2 +\rho^+/2} -\frac{1}{n+1/2 +\rho^+/2t}\right)  \nonumber\\
&&=   \psi\left(\frac{\rho^+ }{2t}+\frac{1}{2}\right) - \psi\left(\frac{\rho^+ }{2}+\frac{1}{2}\right)\nonumber\\
&& \approx -\ln t - \frac{1-t^2}{6\rho^2_+}\,.
\label{transf}
\end{eqnarray} 
The  parameter $ \rho^+$
is large if $T_c\to 0$ and one can use large arguments
 asymptotics of the di-gamma functions. 
Combining Eqs.\,(\ref{self-cons-a}) and (\ref{transf}) we obtain the
self-consistency equation in the form:
      \begin{equation}
\frac{\Psi(1-t^2)}{12\pi T\rho^2_+}  = \sum_{\omega
>0}^{\infty}\left(\frac{\Psi}{\omega^+} +\frac{\Psi\langle\Omega 
\rangle^2}{2t \omega^m_c\omega_c^+\tau_-}-\Big\langle \Omega \, f
\Big\rangle\right)  \,. 
\label{self-cons1}
\end{equation}
 

\subsection{Linearized GL equation and the coherence length}

The GL equations are obtained by utilizing smallness of
$\Delta/\omega$ and of $ {\bm v} {\bm  \Pi}\Delta/\omega^2$ near
$T_c$. Hence, one can use Eqs.\,(\ref{f_1}),
(\ref{f_2}), and (\ref{f_3}) for $f$ and the self-consistency
equation.  For the case of exclusively transport scattering ($\tau_m=\infty$), the GL equations have been derived in Ref.\,\onlinecite{PP}. It is done below taking a finite $\tau_m$. 

To write the self consistency
Eq.\,(\ref{self-cons1}) near $T_c$ one has to express $\langle
\Omega f \rangle$ with the help of Eq.\,(\ref{Eil11}). To this end,
one applies
$\Big\langle \Omega /\omega_+ ...\Big\rangle$ to (\ref{Eil11})  keeping terms up to the order $\delta t$:
\begin{equation}
\Big\langle \Omega \, f\Big\rangle  = \frac{\Psi}
{\omega_+}
+\frac{\langle\Omega \rangle\langle f 
\rangle  }{2\tau_-\omega_+ } -\Big\langle \frac{\langle\Omega 
\rangle  }{2 \omega_+ } {\bm v} {\bm  \Pi}f
\Big\rangle    \, , 
\label{eq25}
\end{equation}
and substitutes the result to Eq.\,(\ref{self-cons1}):
      \begin{equation}
\frac{\Psi\,\delta t }{6\pi T\rho^2_+}  = \sum_{\omega
>0}^{\infty}\left[ \frac{\Psi\langle\Omega 
\rangle^2}{2t \omega _{mc}\omega_{+c}\tau_-}- \frac{ \langle\Omega \rangle \langle f 
\rangle}{2 \omega^+\tau_- } + \Big\langle \frac{\Omega}{2 \omega^+ } \,   {\bm v} {\bm  \Pi}f\Big\rangle  \right]  . 
\label{self-cons4}
\end{equation}
Since we are expanding in powers of $\sqrt{\delta t}$,   the 
distinction between, e.g., $\omega_c$   and  $\omega = \omega_c(1-\delta t) $ is relevant. 

When substituting here $f=f_1+f_2$ of Eqs.\,(\ref{f_1}) and (\ref{f_2}) note that $\langle  \Omega  \,   {\bm v} {\bm  \Pi}\Delta \rangle=0$ because  the angular dependence of $\Omega$ (the symmetry of $\Delta$) has nothing to do with  that of the vector $ {\bm  \Pi}\Delta$. We then  obtain:
      \begin{eqnarray}
\frac{\Psi\,\delta t }{6\pi T\rho^2_+}  &=& \sum_{\omega
>0}^{\infty} \Big[ \frac{ \Psi\langle\Omega 
\rangle^2}{2 \tau_-\omega^2_+ }\left(\frac{ \omega_+^2 
 }{ t \omega^m_c\omega_{+c}}- 1-\frac{1}{2 \tau_-\omega_m}\right)\nonumber\\
&-&\frac{1}{4 \omega^3_{c+}} \Big\langle ({\bm v} {\bm  \Pi})^2\Psi \left(\Omega^2+\frac{\Omega\langle\Omega\rangle}{2 \omega_{cm}\tau_- }\right)\Big\rangle  \Big]  . 
\label{s-cons6}
\end{eqnarray} 
Note that the LHS and the term at the lower line of this 
equation are 
of the order $\delta t^{3/2}$; for this reason  all $\omega$'s in
this term are taken at $T_c$. Besides, the round brackets at the
upper line of the RHS are easily shown to turn zero at $t=1$.
Expanding the bracketed expression in powers of  $\delta t$ and
keeping only the first term one obtains:
      \begin{eqnarray}
  A\, \Psi\,\delta t &=& -B_{ik}  \,  \Pi_i\Pi_k  \Psi \label{AB}\end{eqnarray} 
 with  
  \begin{eqnarray}
   A &=&  \frac{1}{6\pi T_c\rho^2_+} \nonumber\\
   &-&  \frac{ \langle\Omega \rangle^2} {2 \tau_-}  
\sum_{\omega>0}^{\infty}\Big[\frac{
  \omega^2_ + -2\omega\omega_m}{\omega_m \omega_+^3}
-\frac{(2\omega_m+ \omega_+)\omega}{2\tau _- \omega_m^2\omega_+^3
}\Big]  ,\qquad \label{A}\\
 B_{ik}&=& \frac{1}{4} \sum_{\omega>0}^{\infty}\frac{1}{ \omega^3_{ +}} \Big\langle v_iv_k  \left(\Omega^2+\frac{\Omega\langle\Omega\rangle}{2 \omega_{m}\tau_- }\right)\Big\rangle \, , \qquad\qquad
\label{B}
\end{eqnarray} 
where all $\omega$'s are at $T_c$ and the subscript $c$ is omitted. This is, in fact, the linearized anisotropic GL equation 
      \begin{equation}
-(\xi^2)_{ik} \Pi_i\Pi_k  \Psi =\Psi. 
\label{GL-linear}
\end{equation}
with anisotropic coherence length given by 
      \begin{equation}
(\xi^2)_{ik} =B_{ik}/A\delta t\,.
\label{xi-tensor}
\end{equation}

All sums in Eqs.\,(\ref{A}) and (\ref{B}) are expressed in terms of
poly-gamma functions of large parameters $\rho_\pm$. Keeping the
leading terms we obtain:
  \begin{eqnarray}
   A &=&  \frac{1}{6\pi T_c\rho^2_+} - 
 \frac{ \langle\Omega \rangle^2(2\rho_+-\rho^-)} {\pi T_c\rho_-}  
\ln\frac{\rho_+}{2\rho_m}  ,\qquad \label{A1}\\
 B_{ik}&=& \frac{\langle \Omega^2v_iv_k\rangle\tau_+^2}{2\pi
T_c}\nonumber\\ 
&+&\frac{\langle\Omega \rangle\langle \Omega
v_iv_k\rangle\tau_-^2}{2\pi T_c}\left[
\ln\frac{\rho_+}{2\rho_m} - \frac{\rho_-(2 \rho_++\rho_-)}
{2\rho_+^2}\right].\qquad
\label{B1}
\end{eqnarray} 

\subsection{Materials with $\langle\Omega\rangle = 0$ near $T_c$}

This corresponds, e.g.,  to the d-wave symmetry. Within a two-band model for iron-pnictides
 the order parameter has a   $\pm s$ structure, so that $\langle\Delta\rangle \ll |\Delta_{max}|$.\cite{Junhwa}  One then expects the model with $\langle\Omega\rangle = 0$ to hold at least qualitatively for iron-pnictides.


If $\langle\Omega\rangle = 0$, $A$ and $B$ are 
simplified:
      \begin{eqnarray}
     A =  \frac{1}{6\pi T_c\rho^2_+}   \,,\qquad
 B_{ik}= \frac{\langle \Omega^2v_iv_k\rangle\tau_+^2}{2\pi
T_c}\label{B}
\end{eqnarray} 
We then have:
      \begin{equation}
(\xi^2)_{ik} =\frac{3\langle \Omega^2 v_iv_k \rangle\,.
}{4\pi^2T_c^2\delta t }\,\label{xi-d}
\end{equation}

For the  d-wave order parameter and isotropic 2D Fermi surface, $\Omega=\sqrt{2}\cos 2\varphi$ and $\langle \Omega^2 v_x^2\rangle =v^2/2$:
      \begin{equation}
 \xi^2  =\frac{3\hbar^2 v^2}{8\pi^2T_c^2\delta t }\, \,.
\label{x-d-is}
\end{equation}
This result has been obtained in Ref.\,\onlinecite{115} for a clean d-wave with a strongly suppressed $T_c$.

For a uniaxial material, the slope of  the upper critical field along the
$c$ direction near $T_c$ is given by
      \begin{equation}
\frac{dH_{c2,c}}{dT}  = -\frac{2\pi \phi_0 k_B^2 }{ 3\hbar^2\langle
\Omega^2 v_a^2    \rangle}\,T_c  \,  
\label{slope}
\end{equation}
(in common units).
Although the scattering and pair-breaking parameters do not enter this result explicitly, they affect $H_{c2,c}$ and its slope via $T_c(\rho_+)$.  One readily obtains for the other principal direction:
      \begin{equation}
\frac{dH_{c2,ab}}{dT}  = -\frac{2\pi \phi_0 k_B^2 }{3\hbar^2\sqrt{ \langle
\Omega^2 v_a^2    \rangle \langle
\Omega^2 v_c^2    \rangle}}\,T_c  \,.  
\label{slope_ab}
\end{equation}

It is worth recalling that in isotropic materials with the 
standard s-wave order parameter the slope $H_{c2}^\prime \propto
T_c$ in the clean limit (because $H_{c2} \propto 1/\xi^2\propto
T_c^2$) whereas for the dirty case $H_{c2}^\prime$ is $T_c$
independent ($H_{c2} \propto 1/\xi\ell\propto T_c $, $\ell$ is the
mean-free path). The propotionality  $H_{c2}^\prime$ to $T_c $ is a
property of the  AG gapless state. In our case, the result
(\ref{slope}) is obtained  for a strong pair-breaking in materials
with anisotropic order parameter. 
 
Note also  that even without magnetic scatterers,  in   materials with $\langle\Omega\rangle = 0$ and  $\rho^+ \gg 1$,  the
superconductivity becomes ``gapless" in a sense that the {\it total } density of states at the Fermi level is not zero. As in the AG case, if $T_c\to 0$, the superconductivity is  weak  at all temperatures, i.e., $f<<1$ and $g=1-f^2/2=1- \Delta^2/ 2 \omega_+2$ in the whole domain  $0<T<T_c$. 
Then the energy dependence of the total 
density of states $N(\epsilon)=N(0)\,{\rm Re}\,g(\hbar\omega\to
i\epsilon)$ reads:
\begin{eqnarray}
  \frac{N(\epsilon)}{N(0)}= 1-
 2\Delta^2\tau_+^2 \,\frac{1-\eta^2}
{(1+\eta^2)^2} \,,\qquad \eta= 2\tau_+\epsilon .
\label{N(e,gapless)}
\end{eqnarray}
Hence, at zero energy, $N(\epsilon)$ has a non-zero minimum, whereas the maximum of $N(\epsilon)$ is reached at $\epsilon_{m}=   \sqrt{3}/2\tau_{+}$ (not at $\Delta$). Therefore, the ratio of the ``apparent gap" $\epsilon_{m}$ to $T_c$ should vary as $1/T_c$. Since only the total density of states is non-zero, this does not exclude possibility to have gapped and gapless patches on the F-surface.

\section{The specific heat jump} 


Eilenberger equations (\ref{Eil1}) and (\ref{gap1}) in zero field can be
obtained minimizing the functional\cite{E}
\begin{eqnarray}
    {\cal F}&=&N(0)\left[\Psi^2\ln {T\over T_{c0}}
 +2\pi T\sum_{\omega>0} \left(\frac{\Psi^2}{ \hbar \omega
}-\Big\langle I\Big\rangle\right)\right],\qquad\label{Omega}\\
I&=&2\Delta f +2 \omega (g-1)+ \frac{  f\langle f\rangle }{
2\tau^-}+\frac{ (g\langle g\rangle -1)}{2 \tau^+}.\qquad
\label{I}
\end{eqnarray}
 The function $g$ here is an abbreviation for $\sqrt{1-f^2}$. 
Taking account of the self-consistency equation (\ref{gap1}), 
we obtain the energy  difference between the normal and superconducting states:
\begin{eqnarray}
&-&{ F_s-F_n\over 2\pi TN(0)} \\
&=& \sum_{\omega>0} \Big\langle \Delta f +2 \omega (g-1)
+\frac{ f\langle f\rangle }{ 2\tau^-}+\frac{ g\langle g\rangle
-1 }{2 \tau^+}\Big\rangle .\nonumber 
 \label{energydifference}
\end{eqnarray}
One can check that this reduces to the known 
 result for isotropic s-wave cases with or without pair breaking.\cite{Maki} 
This offers a   straightforward way to calculate the 
  the specific heat near $T_c$. The calculation, in general, is tedious
because one has to keep track of  terms up to   $\Delta^4\propto \delta t^2$.
We consider only the case  $\langle \Delta\rangle=0$. 

Up to the forth order in $\Delta$ we have with the help of Eqs.\,(\ref{f_1}) and (\ref{f_3}):
\begin{eqnarray}
f&=&\frac{\Delta}{\omega_+}
 +\frac{\Delta}{2\omega_+^3}\left(\frac{
\langle\Delta^2\rangle}{2\tau_+\omega_+}-\Delta^2\right) ,
\label{f4}\\
g&=&1-\frac{\Delta}{2\omega_+}
+\frac{3\Delta^4}{8\omega_+^4} -\frac{\Delta^2
\langle\Delta^2\rangle}{4\tau_+\omega_+^5},
\label{g4}
\end{eqnarray}
where all $\omega$'s are taken at $T_c$.
Substituting these in the energy difference  we   obtain:
\begin{eqnarray}
-{ F_s-F_n \over 2\pi TN(0)} = \frac{\Psi^4}{4}\sum\left(
\frac{\langle\Omega^4\rangle}{\omega_+^3}-\frac{1}{2\tau_+
\omega_+^4}\right)\,.
\label{FFF}
\end{eqnarray}
For large 
$\rho_+$ one finds:
\begin{eqnarray}
 \sum\left(
\frac{\langle\Omega^4\rangle}{\omega_+^3}-\frac{1}{2\tau_+
\omega_+^4}\right)\approx \frac{(3\langle\Omega^4\rangle
-2)\tau_+^2}{3\pi T }\,.
\label{FFFF}
\end{eqnarray}

To complete the energy evaluation one needs $\Psi(T)$ which is
obtained with the help of the self-consistency equation
(\ref{self-cons4}) and the expression (\ref{f4}) for $f$:
\begin{eqnarray}
 \Psi^2 =\frac{4\pi^2T_c^2(1-t)}{3\langle\Omega^4\rangle-2  } \,.
\label{Psi}
\end{eqnarray}
Thus   the energy difference between the normal and
superconducting states reads:
\begin{eqnarray}
F_n-F_s=    \frac{8\pi^4N(0)\tau_+^2}{3\hbar^2(3\langle\Omega^4\rangle
-2)}\,k_B^4T_c^2(T_c-T)^2\, 
\label{F5}
\end{eqnarray}
in common units.
The specific heat jump at $T_c$ follows:
\begin{eqnarray}
\Delta C=C_s-C_n=
\frac{16\pi^4k_B^4N(0)\tau_+^2}{3\hbar^2(3\langle\Omega^4\rangle -2)}\, T_c
^3\,.
\label{DeltaC}
\end{eqnarray}
Within a weak coupling scheme, this result in a more general form
has been obtained in Ref.\,\onlinecite{Op2}. 

 For the  d-wave state on a cylindrical Fermi surface
 $\Omega=\sqrt{2}\cos^22\phi$ and $\langle\Omega^4\rangle=3/2$ this
gives:
\begin{eqnarray}
\Delta C =  \frac{32\pi^4k_B^4N(0)\tau_+^2}{15\hbar^2} \, T_c ^3\,.
\label{DCd}
\end{eqnarray}

 \section{Discussion}
 
\begin{figure}[b]
\includegraphics[width=8cm]{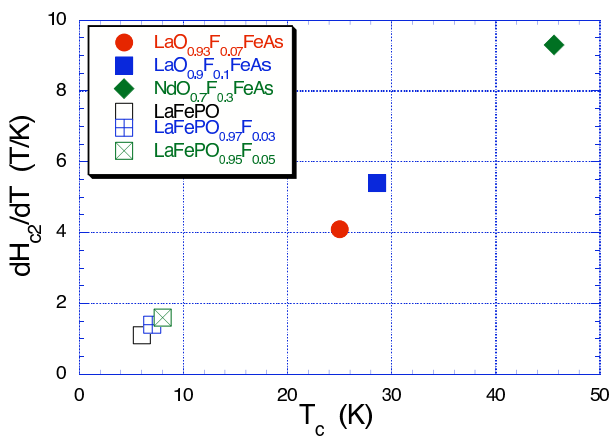}
\caption{(Color online)   The slopes of $H_{c2}(T)$ near $T_c$ (the absolute values) for 
a few 1111  compounds. 
The data for the first three compounds in the legend are taken from   Ref.\,\onlinecite{1111Fuchs}; the remaining three points are taken from  Ref.\,\onlinecite{Suzuki}. The two right-most points are for $H_{c2,ab}^\prime$ of crystalline samples; the rest are for polycrystals, so that all points, in fact, reflect $H_{c2,ab}^\prime$.
}
\label{fig1}
\end{figure}

Figure\,\ref{fig1} is a compilation of data on the slopes
$H_{c2}^\prime$ for 1111 compounds
 with various dopants and, therefore, with various $T_c$'s. An
approximate scaling $H_{c2}^\prime\propto T_c $ is evident
despite the fact that the   compounds examined have 
 $T_c$'s varying from 6 to 46\,K. 
   From this  data  one estimates the slope of
$dH_{c2}^\prime/dT_c $ as
$\approx 0.2 \,$T/K$^2$. Then,   the order of
magnitude of the Fermi velocity follows from $|dH_{c2}^\prime/dT_c|\sim 
 \pi \phi_0 k_B^2/ \hbar^2  v ^2 $ as $v\sim 10^7\,$cm/s, a
reasonable order that can be taken as yet another argument in
favor of the picture presented. 

\begin{figure}[b]
\includegraphics[width=8cm]{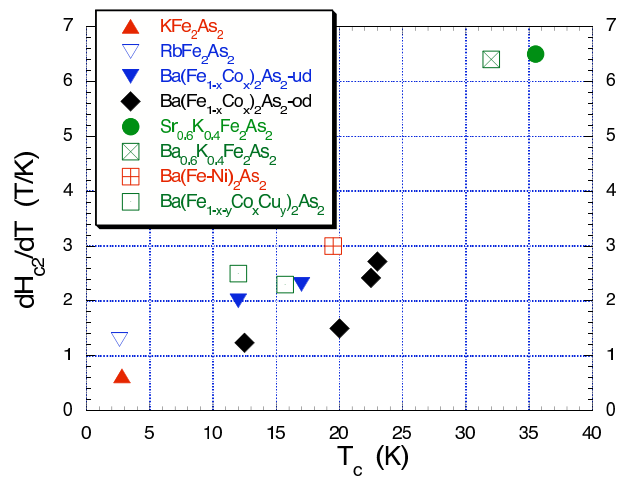}
\caption{(Color online)  The slopes $H_{c2}(T)$ near $T_c$ for 
a few 122 iron-pnictides. The  data   are taken from: 
 RbFe$_2$As$_2$  -- Ref.\,\onlinecite{RbFe2As2}, 
 KFe$_2$As$_2$ -- Ref.\,\onlinecite{KFe2As2},  Ba$_{0.55}$K$_{0.45}$Fe$_2$As$_2$ -- Ref.\,\onlinecite{Ba0.55K0.45Fe2As2}, the underdoped (ud) and overdoped (od) Ba(Fe$_{1-x}$Co$_x$)$_2$Fe$_2$As$_2$ -- Ref.\,\onlinecite{122K}, Ba$_{0.6}$K$_{0.4 }$Fe$_2$As$_2$ -- Ref.\,\onlinecite{Welp}, Sr$_{0.6}$K$_{0.4 }$Fe$_2$As$_2$ -- Ref.\,\onlinecite{Sr0.6K0.4Fe2As2}, Ba(Fe-Ni)$_2$Fe$_2$As$_2$   and Ba(Fe$_{1-x-y}$Co$_x$Cu$_y$)$_2$Fe$_2$As$_2$ -- Ref.\,\onlinecite{Ni}.
}
\label{fig2}
\end{figure}

In Fig.\,\ref{fig2} the data for the 122 family are collected.  The same approximate scaling is seen. 
A considerable scatter of the data points  might
be caused by variety of reasons:  different
criteria in extracting $H_{c2}$ from resistivity data,  
unavoidable uncertainties rooted in sample inhomogeneities in
determination of $T_c$ and the slopes of $H_{c2}(T)$ near $T_c$,     possible
differences in Fermi velocities and the order parameter anisotropies, to name a few. Moreover, the model employing 
only two scattering parameters for  multi-band iron-pnictides is a far-reaching simplification, so that one can expect the model to work  qualitatively at best. 
Nevertheless, the observed scaling seems remarkably robust. One can
take this as evidence in favor of a strong pair-breaking present in
all compounds examined. 
It should be stressed again that for   strongly anisotropic order parameters, $\langle \Delta\rangle\approx 0$, the $T_c$ suppression (or  the pair-breaking, which is the same) is caused by the combined effect of the transport and the spin-flip scattering.


\begin{figure}[htb]
\includegraphics[width=8cm]{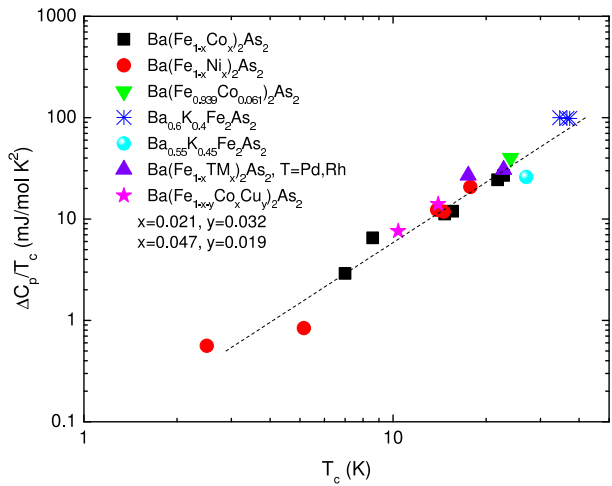}
\caption{(Color online)  The  specific  heat jump versus $T_c$ for 
a few 122 compounds shown on a log-log plot. The dashed line corresponds to  $\Delta
C\propto T_c^3$. Most of the data are from Ref.\,\onlinecite{BNC}; the new data points for mixed Co-Cu doping are shown by stars and taken by the same group, but have not been included in the original publication. }
\label{fig3}
\end{figure}

Figure \ref{fig3} shows the specific heat jump measured in a number of
compounds and reported in Ref.\,\onlinecite{BNC}. The 
``Ames scaling" $\Delta
C\propto T_c^3$ suggested by Bud'ko, Ni, and Canfield is 
evident. Again, it is worth noting that only the combined rate $\rho_+$ enters the
coefficient in front of $T_c^3$ of Eq.\,(\ref{DeltaC}), so that the source of  $T_c$
suppression  is
not necessarily the spin-flip AG pair-breaking. The ever present
transport scattering suppresses $T_c$ as well, provided the order
parameter is strongly anisotropic. This is presumably the case of
iron-pnictides. 

One may wonder why  the scaling $H_{c2}^\prime\propto T_c$
and $\Delta
C\propto T_c^3$ seem to work across the whole class of iron
pnictides for  compounds with different couplings, F-surfaces etc.
Clearly, the source of this scaling should be universal ascross the
phictides family of materials. The pair breaking is offered here as
such an universal source. 

As for the apparent simplicity of the model used one should have in
mind the often overlooked strength of the weak-coupling
scheme: the model is formulated in terms of the measured critical
temperature $T_c$, in which the coupling constants and energy
scales of the ``glue bosons" are incorporated.

 Having succeeded in describing the  ``Ames scalings" just discussed, one can venture 
 to a prediction: according to Eq.\,(\ref{N(e,gapless)}), tunneling experiments are likely to show the ratio of the apparent gap (the maximum position of the total density of states) to $T_c$ varying as $1/T_c$ across the family of iron pnictides.
 

\acknowledgements

Numerous discussions and help of my colleagues S. Bud'ko, Ni Ni, P.
Canfield, J. Schmalian, Junhua Zhang, R. Prozorov, M. Tanatar,  R. Mints, and J. Clem are
appreciated. The work was supported by the US Department of 
Energy, Office of Basic Energy Sciences. 

\newpage
\appendix

\section{Materials with $\langle\Omega\rangle \ne 0$ near $T_c$}

Interestingly enough, the behavior of the $H_{c2}$ slopes as
functions of $T_c$ turns out different if 
$\langle\Omega\rangle \ne 0$. 
To see this, consider the
expression for the coefficient $A$ of Eq.\,(\ref{A1}). In terms of
scattering times, it reads:
  \begin{eqnarray}
   A = \frac{2\pi T_c\tau_+^2}{3} - 
 \frac{ \langle\Omega \rangle^2 } {\pi T_c
}\left(\frac{2\tau_-}{\tau_+}-1\right)  
\ln\frac{\tau_-}{2\tau_+} \,. \label{A2} 
\end{eqnarray} 
Since all $\tau$'s are finite near the critical point where 
$T_c\to 0$, the  term $\propto \langle\Omega \rangle^2$ is leading.
Consider, e.g.,  a usual situation $\tau<<\tau_m$:
 \begin{eqnarray}
   A \approx   
 \frac{ \langle\Omega \rangle^2 \ln 2} {\pi T_c
} \,. \label{A3} 
\end{eqnarray} 
After simple algebra one obtains   the slope of 
$H_{c2,c}$ at $T_c$:
      \begin{equation}
\frac{dH_{c2,c}}{dT}  = -\frac{ \phi_0  }{2\pi\tau^2 T_c}\, 
\frac{\langle\Omega\rangle^2\ln 4}{\langle v_a^2\Omega^2\rangle
+\langle\Omega\rangle \langle v_a^2\Omega \rangle 
\ln(2\tau_m/\tau e^{ 3/2})   
}. 
\label{slope1}
\end{equation}
Thus, the slopes $H_{c2}^\prime \propto 1/T_c$, the dependence opposite to that of the
 case  $\langle\Omega\rangle = 0$.


\begin{thebibliography}{99}

\bibitem{BNC} S.L. Bud'ko, Ni Ni, and P.C. Canfield, \prb {\bf 79},
220516(R) (2009).

\bibitem{Z}J. Zaanen, cond-mat: 0908.0033.

\bibitem{122K}
N. Ni,  M.E. Tillman,  J.-Q. Yan,  A. Kracher, 
S.T. Hannahs,  S.L. BudÕko,  and P.C. Canfield, cond-matt/0811.1767. 


\bibitem{AG}A. A. Abrikosov and L. P. Gor'kov, Zh. Eksp. Teor. Fiz. 
{\bf 39}, 1781 (1060) [Sov. Phys. JETP, {\bf  12}, 1243 (1961)].

\bibitem{Mazin-Schm}I.I. Mazin, J. Schmalian,  Phys. C: Supercond,  {\bf 469}, 614 (2009).


\bibitem{Junhwa} Junhua Zhang, R. Sknepnek, R. M. Fernandes, and J. Schmalian, \prb {\bf 79}, 220502 (2009). 


\bibitem{E}G. Eilenberger, Z. Phys. {\bf  214}, 195 (1968).

  \bibitem{Maki}K. Maki in {\it Superconductivity} ed by R. D. Parks,
Marcel Dekker, New York, 1969, v.2, p.1035.

\bibitem{Kad} D. Markowitz, L.P. Kadanoff, Phys. Rev.\,{\bf 131}, 363
(1963).


 \bibitem{Openov}L. A. Openov, JETP Lett. {\bf 66}, 661 (1997).


\bibitem {115}V.G. Kogan, R. Prozorov, and C. Petrovic,  J. Phys.: Condens. Matter {\bf 21}, 102204 (2009). 

 \bibitem{Op2}L. A. Openov, \prb. {\bf 69}, 224516 (2004).

 \bibitem{PP}S.V. Pokrovsky and V.L. Pokrovsky, \prb {\bf 54}, 13275 (1996)

\bibitem{1111Fuchs}  G. Fuchs, S-L. Drechsler, N. Kozlova, M. Bartkowiak,
J.E. Hamann-Borrero, G. Behr, K. Nenkov, H-H. Klauss,
H. Maeter, A. Amato, H. Luetkens, A. Kwadrin,
R. Khasanov, J. Freudenberger, A. Koehler, M. Knupfer,
E. Arushanov, H. Rosner, B. Buechner, and L. Schultz, cond-matt/0902.3498 (2009).

\bibitem{Suzuki}S. Suzuki, S. Miyasaka, S. Tajima, T. Kida, and
M. Hagiwara, cond-matt/0910.1711.

\bibitem{Ba0.55K0.45Fe2As2}M. M. Altarawneh, K. Collar, C. H. Mielke, 
N. Ni, S. L. BudÕko, and P. C. Canfield, \prb {\bf 78}, 220505R (2008).

\bibitem{RbFe2As2}Z. Bukowski, S. Weyeneth, R. Puzniak, J. Karpinski, B. Batlogg, cond-matt/09092740.

\bibitem{KFe2As2} T. Terashima, M. Kimata, H. Satsukawa, A. Harada, K. Hazama, S. Uji, H. Harima, Gen-Fe Chen,Jian-Jin Luo, and Nan-Lin Wang, Journ. Phys. Soc. Japan, {\bf 78}, 063702 (2009). 


 \bibitem{Welp}U. Welp, G. Mub, R. Xie, A.E. Koshelev, W.K. Kwok, H.Q. Luo, Z.S. Wang, P.  Cheng, L. Fang,
 C. Ren, H.-H. Wen, Phys. C, {\bf 469}, 575 (2009).
 
 \bibitem{Sr0.6K0.4Fe2As2}        G. F. Chen, Z. Li, J. Dong, G. Li, W. Z. Hu, X. D.
Zhang, X. H. Song, P. Zheng, N. L. Wang, and J. L. Luo, cond-matt/0806.2648. 


 \bibitem{Ni}Ni Ni, private communication.

\end{thebibliography}
\end{document}